\documentclass[11pt]{article}
\usepackage[margin=1in]{geometry}
\usepackage{times}
\usepackage{titlesec}
\usepackage{hyperref}
\usepackage{graphicx}
\usepackage{float}
\usepackage{amsmath}
\usepackage{caption}
\usepackage{fancyhdr}
\usepackage{multicol}
\usepackage{tcolorbox}
\usepackage{fancyvrb}
\usepackage{caption} 
\usepackage{listings}
\usepackage{xcolor}

\definecolor{codebg}{RGB}{248,248,248}

\titleformat{\section}{\large\bfseries}{\thesection}{1em}{}
\titleformat{\subsection}{\normalsize\bfseries}{\thesubsection}{1em}{}

\hypersetup{
    pdftitle={VLAI: A RoBERTa-Based Model for Automated Vulnerability Severity Classification},
    pdfauthor={Cédric Bonhomme, Alexandre Dulaunoy},
    pdfsubject={2025 Vulnerability Forecasting Technical Colloquium, Cambridge, United Kingdom},
    pdfkeywords={CVE, Vulnerability, NLP, AI, Classification, RoBERTa},
    colorlinks=true, linkcolor=blue, citecolor=blue, urlcolor=blue
}

\title{\textbf{VLAI: A RoBERTa-Based Model for Automated Vulnerability Severity Classification}}
\author{
    Cédric Bonhomme \\
    \textit{Computer Incident Response Center Luxembourg} \\
    \href{mailto:cedric.bonhomme@circl.lu}{cedric.bonhomme@circl.lu} [\href{https://openpgp.circl.lu/pks/lookup?op=get\&search=0xA1CB94DE57B7A70D}{57B7 A70D}]
    \and
    Alexandre Dulaunoy \\
    \textit{Computer Incident Response Center Luxembourg} \\
    \href{mailto:alexandre.dulaunoy@circl.lu}{alexandre.dulaunoy@circl.lu} [\href{https://pgp.circl.lu/pks/lookup?op=get\&search=0x3b12dcc282fa29312f5b709a09e2cd4944e6cbcd}{44E6 CBCD}]
}
\date{2025-09-25}

\begin{document}
\maketitle

\begin{abstract}
This paper presents VLAI, a transformer-based model that predicts software vulnerability severity levels
directly from text descriptions.
Built on RoBERTa, VLAI is fine-tuned on over 600,000 real-world vulnerabilities and achieves over
82\% accuracy in predicting severity categories, enabling faster and more consistent triage ahead of manual CVSS scoring.
The model and dataset are open-source and integrated into the Vulnerability-Lookup service.
\end{abstract}

\vspace{1em}
\noindent\textbf{Topics:} \textit{Measuring vulnerabilities, Exploits or exploitation}, \textit{Decision science of vulnerability management}

\section{Introduction}
Thousands of new software vulnerabilities are disclosed every year, often initially with only a brief textual
description and without an official severity score. Security experts later analyze these vulnerabilities
and assign severity ratings using standards like the Common Vulnerability Scoring System (CVSS).
However, this manual assessment process can take days, leaving a critical gap where defenders must
prioritize vulnerabilities without clear guidance. To bridge this gap, we present VLAI (Vulnerability
Lookup AI) – an NLP model that predicts a vulnerability’s severity directly from its description, before any
official score is available. Our approach leverages a fine-tuned RoBERTa transformer~\cite{DBLP:journals/corr/abs-1907-11692} to classify
vulnerability descriptions into severity categories, enabling security analysts to obtain an immediate
estimated severity (the “VLAI score”) based solely on the description. The entire solution is open-source
and integrated into the Vulnerability-Lookup service, providing the community with timely severity
estimates and a continuously improving model.

\section{Related Work}
Early efforts to automate vulnerability severity estimation used classical NLP and machine learning
techniques. For example, statistical models have been trained on vulnerability descriptions to predict CVSS
metrics or scores with varying success. Recent advances in deep learning have enabled more accurate
predictions: Shahid et al. (2021)~\cite{shahid2021cvssbertexplainablenaturallanguage} used BERT-based classifiers (in a system called CVSS-BERT) to determine the
full CVSS vector from a description, and reported that the computed severity scores were very close to the
actual scores assigned by human experts.
Industry practitioners have also shown interest in this
problem; for instance, PRIOn developed NLP models to predict CVSS base scores and vectors from text,
citing the delay in official scoring and the need for faster risk assessment. Our work follows this line
of research by fine-tuning a Transformer-based model on an even larger consolidated dataset of
vulnerabilities, and is, to our knowledge, one of the first open and continuously updated deployments of
such a model integrated into a public vulnerability lookup service.

Recent research has also explored the use of semantic similarity techniques for structured vulnerability classification.
Kota et al. \cite{KOTA20241167} proposed a novel approach to predict CWE identifiers from CVE descriptions using a
cross-encoder architecture. Their method leverages the hierarchical structure of the MITRE CWE taxonomy and trains
separate cross-encoder models for each layer of the CWE tree. By combining these models with a binary classifier,
they achieved an overall test accuracy of 72.1\% and a macro-averaged F1 score of 0.735 across nearly 14,000 CVE entries.
This work demonstrates the potential of semantic text similarity and layered classification strategies for automating
weakness identification, complementing efforts focused on severity classification.

\section{System Architecture and Data Pipeline}

\begin{figure}[H]
    \centering
    \includegraphics[width=\linewidth]{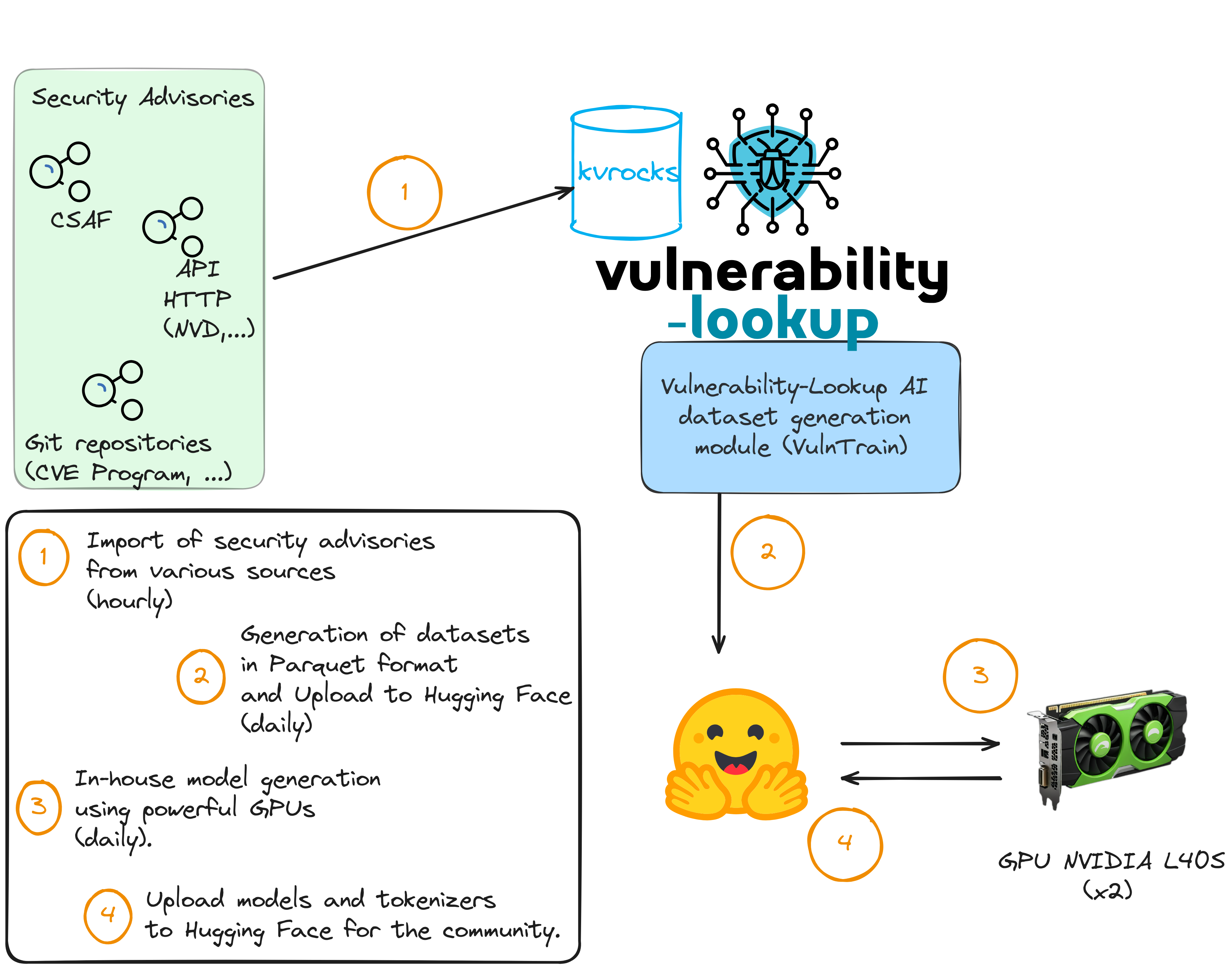}
    \caption{Automated pipeline for dataset creation and model training in the Vulnerability-Lookup AI system.}
    \label{fig:ai-workflow}
\end{figure}

A more detailed presentation of the Vulnerability-Lookup internal architecture can be found in \cite{bonhomme-vulnerability-lookup-valencia-2024}.

The pipeline consists of four main stages, executed on a regular schedule to keep the model up-to-date:
\begin{enumerate}
  \item \textbf{Data Collection:} Security advisories from multiple sources are imported into the Vulnerability-
	  Lookup platform~/cite{VulnerabilityLookup} on an ongoing basis (hourly). These sources include official feeds and
repositories such as the CVE Program, NVD APIs, GitHub Security Advisories, PyPI security advisories,
and vendor CSAF feeds (details in Dataset section). All collected advisories are stored in a
centralized database (using a KV storage, kvrocks). 
  \item \textbf{Dataset Generation:} Each day, a snapshot of the accumulated vulnerability data is transformed into
a training dataset for the model. We extract each vulnerability’s unique ID, description text,
associated CPEs (affected product identifiers), and any available CVSS scores (v2, v3.0, v3.1, etc.). This
dataset is saved in a standardized format (Parquet) and uploaded to the Hugging Face Hub (as the
CIRCL/vulnerability-scores dataset) for transparency and community access. The dataset
currently contains ~610k entries split into a training set (~550k) and test set (~60k), and it
grows as new vulnerabilities are disclosed.
  \item \textbf{Model Training:} Using the freshly built dataset, the severity classification model is retrained (fine-
tuned) daily on our GPU servers. We leverage the open-source VulnTrain tool to orchestrate fine-
tuning of a RoBERTa-base model~\cite{DBLP:journals/corr/abs-1907-11692} on this data. The training process (detailed in Section Model
and Training) typically runs for 5 epochs on two NVIDIA L40S GPUs, completing in approximately 6
hours for the full dataset. After training, the updated model and tokenizer artifacts are saved. 
  \item \textbf{Deployment:} The newly trained model checkpoint is automatically uploaded to the Hugging Face
  Model Hub under the organization’s repository ~\cite{huggingfaceCIRCLvulnerabilityseverityclassificationrobertabaseHugging}.
 This allows the community to access the latest model version immediately. The
model is also integrated into the online Vulnerability-Lookup service, so that users querying a
vulnerability with no known CVSS will see the VLAI-predicted severity alongside other information. This continuous cycle ensures that both the data and the model keep pace with the evolving
vulnerability landscape. 
\end{enumerate}

\section{Dataset}

The CIRCL/vulnerability-scores dataset~\cite{ossbaseDatasetsVLAI}~\cite{circldatasetvulnscore} is a comprehensive collection of vulnerability descriptions and
their associated severity scores. It compiles data from multiple respected sources in the security
community, providing a rich and diverse training corpus.
The current sources are: CVE Program (enriched with data from \textit{vulnrichment} and Fraunhofer FKIE), GitHub Security Advisories,
PySec advisories, CSAF feeds from Red Hat, CISCO and CISA. All these sources are configured by default in Vulnerability-Lookup.

The sources of data include (but are not limited to):
\begin{itemize}
	\item CVE Program (NVD): The official MITRE CVE list and NIST’s National Vulnerability Database entries,
		enriched with additional data from projects like vulnrichment and Fraunhofer FKIE~\cite{circlVLFKIE}. 
		These records supply standardized descriptions and CVSS scores assigned by NVD analysts if present.
	\item GitHub Security Advisories (GHSA): Vulnerability advisories from the GitHub Advisory Database: covering security issues in open-source packages (often with CVE IDs and sometimes CVSS v3
scores provided by maintainers or GitHub).
	\item PyPI Security Advisories: Reports of Python package vulnerabilities (PyPI), which contribute
additional descriptions of software flaws not always captured in NVD.
	\item CSAF Feeds: Machine-readable security advisories in CSAF format from vendors such as Red Hat and
Cisco, which include detailed vulnerability information and vendor-assigned severities.
\end{itemize}

Each entry in the dataset~\cite{ossbaseDatasetsVLAI}~\cite{circldatasetvulnscore} represents a single vulnerability and includes several fields: a unique identifier
(e.g. CVE-IDs, GHSA IDs, vendor bulletin IDs), a short title, the full textual description of the
vulnerability, a list of associated CPE names (affected product identifiers), and the CVSS base scores in
various versions

If multiple CVSS versions exist for a given record (e.g., both v2 and v3), all are recorded. Not all vulnerabilities have an official score at first;
many advisories (especially from third-party sources or very recent disclosures) have a non-defined value in these score fields, indicating no score was
assigned yet. This reinforces the need for our model to fill in severity estimates.

The dataset is updated daily alongside the vulnerability feeds. This means new CVEs and advisories
are continually appended. Over time, as previously unscored vulnerabilities receive official scores (e.g., NVD
analysts eventually score a CVE), those entries move from the "unlabeled" pool into the labeled pool for
future training cycles. The daily refresh and publishing of the dataset on Hugging Face Hub ensures
transparency and enables others to use or analyze this data. In summary, the CIRCL/vulnerability-scores
dataset~\cite{ossbaseDatasetsVLAI}~\cite{circldatasetvulnscore} provides a large-scale, diverse, and up-to-date foundation for training the severity prediction model.

\section{Model and Training}
The model uses RoBERTa-base~\cite{DBLP:journals/corr/abs-1907-11692} with a softmax classification head. Training~\cite{VulnTrain} uses 512-token max sequence length, AdamW optimizer, and 5 epochs. With a batch size of 16, training on 2× L40S GPUs takes about 6 hours. Final accuracy on the test set is 82.8\%.

We chose RoBERTa for its robust language understanding; vulnerability descriptions often contain technical terms and context (e.g., buffer overflow,
RCE, privilege escalation) that a large pre-trained model can interpret and relate to severity implications.
Using the Hugging Face Transformers library, we add a classification head on RoBERTa (a linear layer with
softmax) to predict one of four severity categories.

Training procedure: The model is trained~\cite{VulnTrain} using the prepared dataset described above. We feed in the
vulnerability description text as input to the model. (In preliminary experiments, we found that including
the short title or CPE fields did not significantly improve performance beyond what the description provides,
likely because the description already contains the key details about impact and context.) Each input text is
truncated or padded to a maximum sequence length (we use 512 tokens to cover most descriptions) and
encoded with RoBERTa’s byte-pair tokenizer. The corresponding label is the severity class derived from the
CVSS score. We fine-tune the model using a standard cross-entropy loss on this 4-class classification task.

Training is done with the AdamW optimizer (weight decay regularization) and a linear learning rate
scheduler (with warm-up). We used the following hyperparameters based on common defaults and some
tuning: a learning rate of $3e-5$, batch size of 16, and 5 training epochs.

These settings were found to work well given the dataset size and available
computing resources based on empirical results. Each daily training run consumes on the order of 6–7 hours on 2× NVIDIA L40S GPUs
(48 GB each of VRAM) to process $\approx$550k training examples over 5 epoch. Despite the large dataset, RoBERTa
converges effectively within five epochs, as we observed the validation loss stabilizing and accuracy
plateauing by epoch 4–5.

Inference usage: The trained model (dubbed vulnerability-severity-classification-roberta-base) is
available on Hugging Face Hub for inference. To use it, one can simply load the model and tokenizer and
classify new descriptions. For example, using the snippet from our documentation:
\lstset{
  language=Python,
  basicstyle=\ttfamily\small,
  keywordstyle=\color{blue},
  commentstyle=\color{green!50!black},
  stringstyle=\color{red!70!black},
  frame=single,
  breaklines=true,
  tabsize=2,
  showstringspaces=false,
}

\begin{samepage}
\begin{lstlisting}[caption={Predicting vulnerability severity with a fine-tuned RoBERTa model},
                   label={lst:severity}]
from transformers import AutoModelForSequenceClassification, AutoTokenizer
labels = ["low", "medium", "high", "critical"]  # defined severity classes
model_name = "CIRCL/vulnerability-severity-classification-roberta-base"

tokenizer = AutoTokenizer.from_pretrained(model_name)
model = AutoModelForSequenceClassification.from_pretrained(model_name)
model.eval()

# Example vulnerability description:
text = ("A buffer overflow in XYZ software allows remote code execution "
        "with root privileges.")
inputs = tokenizer(text, return_tensors="pt", truncation=True, padding=True)
outputs = model(**inputs)

pred_scores     = outputs.logits.softmax(dim=-1)
predicted_class = pred_scores.argmax(dim=-1).item()
print("Predicted severity:", labels[predicted_class])
\end{lstlisting}
\end{samepage}

This would output a severity prediction (for the example above, one would expect it to predict “critical” due
to the mention of remote code execution). In fact, using a real example: “...allows an attacker to execute
arbitrary code via ...”, the model indeed produces a very high probability for the critical class (98\%
confidence).

Such predictions can be obtained instantly for any new vulnerability description, making
the model a valuable assistant for security analysts.

\section{Evaluation and Results}

To demonstrate the model’s usefulness in practice, we conducted an
evaluation on unscored vulnerabilities. We took a set of recent vulnerabilities that, at the time of their
disclosure, had no CVSS score. We used our model to predict severity for each based
solely on the description. Later, once the vendors or NVD released official CVSS scores for these issues
(serving as ground truth), we compared our predictions to the actual outcomes. In this experiment, the
model’s predicted severity matched the eventual official severity about 85\% of the time.

In other words, for a large majority of these cases, the VLAI model~\cite{computer_incident_response_center_luxembourg_2025} gave the same severity category (e.g., “High”) that the
human analysts later assigned. This is a powerful result: it shows that our system can effectively anticipate
the severity of new vulnerabilities in that critical early period before expert analysis is available.

In the $\approx$15\% of cases where the model did not match, we observed that the predictions were usually off by one
category (for instance, predicting High when it was eventually rated Critical, or Medium vs Low). Rarely did
the model completely miss (e.g., predicting Low for an eventual Critical). This gives us confidence that even
when not perfect, the model’s output can still be useful to approximate risk (often erring on the side of
caution by slightly over-predicting severity, which is preferable in a security context).

\subsection{Real-World Scenario}
In evaluation of recent vulnerabilities without initial CVSS scores, VLAI predictions matched eventual CVSS categories 85\% of the time.

We also note the model’s predictions are fast – inference takes only a fraction of a second per sample on a
GPU, or a few seconds on CPU – which means this approach can be used to automate triage at scale. For
example, if hundreds of new advisories come in a day, the model can instantly label them, helping security
teams prioritize which issues to address first. This complements the slower, manual scoring process: the
model provides an immediate “provisional” severity that can later be confirmed or adjusted by experts~\cite{shahid2021cvssbertexplainablenaturallanguage}.

The integration of our model into the online Vulnerability-Lookup service~\cite{VulnerabilityLookup} means users are already
benefitting from these AI-generated severity scores in real time. Overall, the evaluation confirms that our
RoBERTa-based classifier performs robustly and adds tangible value in vulnerability management
workflows.

\subsection{Inference Integration within Vulnerability-Lookup}

\begin{figure}[H]
    \centering
    \includegraphics[width=\linewidth]{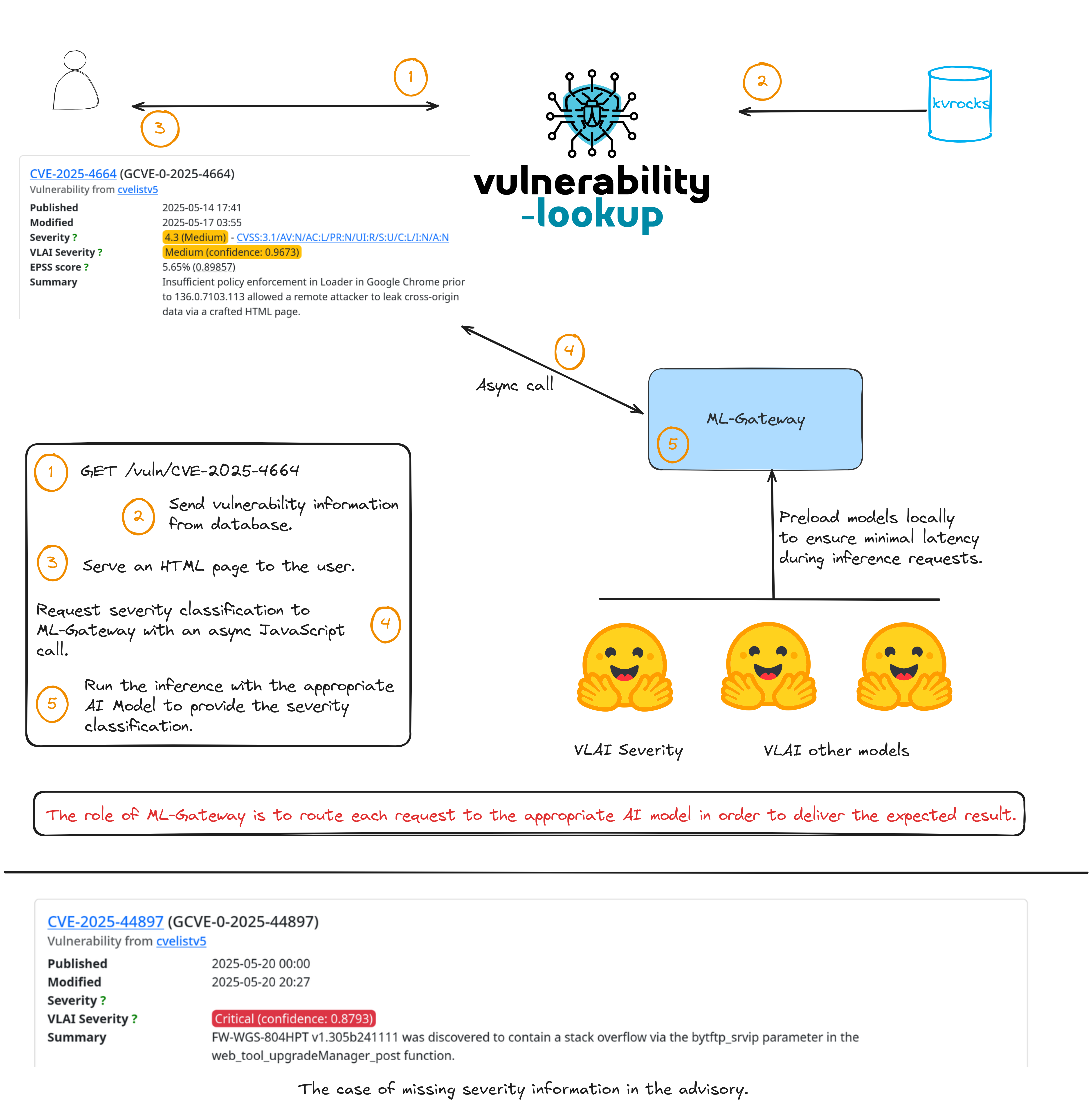}
    \caption{ML-Gateway.}
    \label{fig:ml-gateway}
\end{figure}

ML-Gateway is a lightweight, FastAPI-based server designed to serve pre-trained NLP models locally through a RESTful API interface.
At startup, the gateway loads all configured models into memory, enabling low-latency inference without reliance on external services.

All inference requests are handled entirely on our infrastructure, ensuring that no input data is transmitted to third-party platforms
such as Hugging Face. While the Hugging Face Hub is used for publishing models and datasets to foster openness and reproducibility,
inference remains strictly local to preserve privacy and performance.

Functioning as a streamlined model-serving layer, ML-Gateway simplifies the integration of multiple AI models into the
Vulnerability-Lookup platform without introducing architectural complexity.

Each model is exposed via distinct HTTP endpoints, with auto-generated OpenAPI documentation detailing available routes,
expected inputs, and example responses—facilitating seamless adoption by both internal and external tools.

\section{Conclusion}

We have presented the design and implementation of VLAI, an automated vulnerability severity scoring
model built on NLP techniques. By fine-tuning a RoBERTa transformer on a large, up-to-date dataset of
vulnerability descriptions and scores, we achieved accurate classification of vulnerabilities into standard
severity categories. Our open-source model and dataset, updated on a daily cycle, demonstrate how AI can
augment the vulnerability assessment process by providing immediate severity estimates when no official
score is yet available~\cite{ossbaseDatasetsVLAI}. This assists analysts and organizations in prioritizing patches and mitigations for
critical issues without waiting days or weeks for human triage.

\subsection{Impact}

The VLAI model is integrated into the CIRCL Vulnerability-Lookup platform, making AI-driven
severity insights accessible to the community at no cost. The model’s $\approx$83\% accuracy on historical data and
$\approx$85\% agreement with eventual expert ratings in live tests indicate that it can serve as a reliable “second
opinion” or preliminary risk indicator. As the model is continuously retrained with new data, we expect its
performance to further improve over time or adapt to new trends in vulnerabilities.

\subsection{Risk and Ethical Consideration}

While VLAI provides valuable assistance in estimating the severity of vulnerabilities from textual descriptions, its usage is not without risks and limitations.

One critical concern is the potential for adversarial manipulation. Since the model relies solely on the language of the vulnerability description to infer severity, there is a risk that vendors or vulnerability reporters may intentionally downplay or obfuscate key technical details in the description to influence the predicted score. For example, omitting terms like "remote code execution" or replacing precise impact language with vague wording could lead to a lower severity classification by the model. This presents a misalignment risk where the model’s predictions may not reflect the actual technical impact if the input data is selectively crafted or sanitized.

Additionally, like any AI model, VLAI may reflect biases present in its training data. If certain types of vulnerabilities are historically underrepresented or misclassified in public datasets, the model may generalize poorly or exhibit skewed predictions in edge cases or for emerging vulnerability types.

Therefore, we emphasize that VLAI should be used as a supplementary tool~\cite{circlAIStrategy}, not as a replacement for expert analysis. It is intended to support triage and prioritization in early-stage assessments, particularly when no formal scoring is yet available. Analysts should remain vigilant for potential discrepancies and treat model outputs as provisional guidance rather than authoritative classifications.

\subsection{Future Work}

There are several avenues to enhance this work. One direction is to incorporate
explainability into the model’s predictions – e.g. using attention weights or input saliency methods to
highlight which words in the description most influenced the predicted severity (similar to the approach in
CVSS-BERT for explaining CVSS metric prediction ~\cite{shahid2021cvssbertexplainablenaturallanguage}).

This could increase user trust in the AI recommendations by showing rationales (for instance, highlighting phrases like "execute arbitrary code" for
a Critical prediction). Another improvement would be expanding the model to predict the full CVSS vector or
score value, not just the severity category. This is a more granular task, but techniques from multi-label
classification or regression could be applied, potentially training separate sub-models for each CVSS metric
as done in prior research ~\cite{shahid2021cvssbertexplainablenaturallanguage}.

Beyond vulnerability assessment, the methodology behind VLAI can be extended to other cybersecurity contexts—particularly structured threat intelligence platforms like MISP~\cite{wagner2016misp}. MISP stores rich contextual information in objects and attributes, many of which contain free-text fields written by analysts, including incident summaries, threat actor descriptions, malware characteristics, TTPs, and analytical assessments.

Future research could explore adapting the VLAI model, or training a variant of it, to analyze and classify such textual fields within MISP. This could enable automated tagging of incident severity, threat confidence levels, or detection prioritization based on narrative content. For instance, a transformer model could learn to infer the likely impact or criticality of an event based on observed behaviors or IOCs, or help identify high-risk threats by comparing descriptions to historical incident patterns.

While the current VLAI model is trained on English-language vulnerability descriptions, the underlying methodology can be extended to support other languages by leveraging multilingual or language-specific RoBERTa variants~\cite{cui-etal-2020-revisiting}. For example, models such as Chinese RoBERTa (Chinese-RoBERTa-wwm-ext) or multilingual XLM-RoBERTa could be fine-tuned on vulnerability data written in Chinese or other non-English sources, enabling similar severity classification capabilities across different linguistic and regional contexts. This would be particularly beneficial for national CSIRTs, regional vulnerability databases, and organizations operating in multilingual environments where English descriptions are not always available. Such an extension would help democratize automated risk assessment tools and improve early vulnerability triage in underrepresented language domains.

\section*{References}

\bibliography{mybib}{}

\begin{thebibliography}{10}

\bibitem{VulnTrain}
C{\'e}dric Bonhomme.
\newblock vulnerability-lookup/{VulnTrain}.
\newblock \url{https://github.com/vulnerability-lookup/VulnTrain}, jul 2025.

\bibitem{bonhomme-vulnerability-lookup-valencia-2024}
Bonhomme C{\'e}dric.
\newblock Vulnerability-lookup - an open source tool to support cvd processes.
\newblock In {\em Valencia 2024 UNDP/UNICC/FIRST Technical Colloquium}, October
  2024.

\bibitem{VulnerabilityLookup}
CIRCL.
\newblock Vulnerability-lookup official website.
\newblock \url{https://vulnerability-lookup.org/}, 2024.
\newblock [Accessed 04-07-2025].

\bibitem{ossbaseDatasetsVLAI}
CIRCL.
\newblock Ai datasets and vlai model --- discourse.ossbase.org.
\newblock \url{https://discourse.ossbase.org/t/ai-datasets-and-vlai-model/105},
  2025.
\newblock [Accessed 04-07-2025].

\bibitem{circlAIStrategy}
CIRCL.
\newblock Circl.lu - ai strategy.
\newblock \url{https://www.circl.lu/pub/ai-strategy/}, 2025.
\newblock [Accessed 04-07-2025].

\bibitem{huggingfaceCIRCLvulnerabilityseverityclassificationrobertabaseHugging}
CIRCL.
\newblock Circl/vulnerability-severity-classification-roberta-base · hugging
  face --- huggingface.co.
\newblock
  \url{https://huggingface.co/CIRCL/vulnerability-severity-classification-roberta-base},
  2025.
\newblock [Accessed 04-07-2025].

\bibitem{circlVLFKIE}
CIRCL.
\newblock Recent vulnerabilities - vulnerability-lookup -
  vulnerability.circl.lu - fkie source.
\newblock \url{https://vulnerability.circl.lu/recent#fkie_nvd}, 2025.
\newblock [Accessed 04-07-2025].

\bibitem{cui-etal-2020-revisiting}
Yiming Cui, Wanxiang Che, Ting Liu, Bing Qin, Shijin Wang, and Guoping Hu.
\newblock Revisiting pre-trained models for {C}hinese natural language
  processing.
\newblock In {\em Proceedings of the 2020 Conference on Empirical Methods in
  Natural Language Processing: Findings}, pages 657--668, Online, November
  2020. Association for Computational Linguistics.

\bibitem{KOTA20241167}
Kethan Kota, Manjunatha A, and Sree~Vivek S.
\newblock Cwe prediction using cve description - the semantic similarity
  approach.
\newblock volume 235, pages 1167--1178, 2024.
\newblock International Conference on Machine Learning and Data Engineering
  (ICMLDE 2023).

\bibitem{DBLP:journals/corr/abs-1907-11692}
Yinhan Liu, Myle Ott, Naman Goyal, Jingfei Du, Mandar Joshi, Danqi Chen, Omer
  Levy, Mike Lewis, Luke Zettlemoyer, and Veselin Stoyanov.
\newblock Roberta: {A} robustly optimized {BERT} pretraining approach.
\newblock {\em CoRR}, abs/1907.11692, 2019.

\bibitem{circldatasetvulnscore}
Computer Incident Response~Center Luxembourg.
\newblock vulnerability-scores (revision 6764823).
\newblock \url{https://huggingface.co/datasets/CIRCL/vulnerability-scores},
  2025.

\bibitem{computer_incident_response_center_luxembourg_2025}
Computer Incident Response~Center Luxembourg.
\newblock vulnerability-severity-classification-roberta-base (revision
  2100b2d).
\newblock
  \url{https://huggingface.co/CIRCL/vulnerability-severity-classification-roberta-base},
  2025.

\bibitem{shahid2021cvssbertexplainablenaturallanguage}
Mustafizur Shahid and Hervé Debar.
\newblock Cvss-bert: Explainable natural language processing to determine the
  severity of a computer security vulnerability from its description, 2021.

\bibitem{wagner2016misp}
Cynthia Wagner, Alexandre Dulaunoy, G{\'e}rard Wagener, and Andras Iklody.
\newblock Misp: The design and implementation of a collaborative threat
  intelligence sharing platform.
\newblock In {\em Proceedings of the 2016 ACM on Workshop on Information
  Sharing and Collaborative Security}, pages 49--56. ACM, 2016.

\end{thebibliography}
\bibliographystyle{plain}
\end{document}